\documentclass[reprint,amsmath,amssymb,aps,pre,showkeys,showpacs,floats,nofootinbib]{revtex4-1}

\usepackage{graphics}
\usepackage{graphicx}
\usepackage{wrapfig}
\usepackage[center,normalsize]{subfigure}
\usepackage{xspace}
\usepackage{url}
\usepackage{pifont,epsfig}
\usepackage{color}
\usepackage{bm}
\usepackage{textcomp}
\usepackage{mathrsfs}
\usepackage{ctable}
\usepackage{pstricks,pst-node,pst-plot,pst-coil,pst-text,multido,pst-3d,pst-tree,pst-grad}

\definecolor{dgreyblue}{rgb}{0.26,0.3,0.46}             

\newcommand{\cD}{\mathcal{D}}
\newcommand{\R}{{\mathbb R}}  

\renewcommand{\text}[1]{\hbox{\rm \ #1\ \/}}

\newcommand{\be}[1]{\begin{equation}\label{#1}}
\newcommand{\ee}{\end{equation}}
\newcommand{\beqn}{\begin{eqnarray*}}
\newcommand{\eeqn}{\end{eqnarray*}}
\newcommand{\beq}{\begin{eqnarray}}
\newcommand{\eeq}{\end{eqnarray}}
\newcommand{\ben}{\begin{enumerate}}
\newcommand{\een}{\end{enumerate}}
\newcommand{\bi}{\begin{itemize}}
\newcommand{\ei}{\end{itemize}}

\newcommand{\IE}{{\em i.e.}\xspace}
\newcommand{\tx}{^{\rm th}}

\newtheorem{theorem}{Theorem}
\newtheorem{lemma}[theorem]{Lemma}

\newtheorem{definition}[theorem]{Definition}

\newenvironment{proof-sketch}{{\noindent\bf Sketch of Proof.\ }}{\hfill{\Pisymbol{pzd}{113}}\vspace{0.1in}}

\newcommand{\NP}{\mathsf{NP}}
\renewcommand{\deg}{\mathsf{d}}

\newcommand{\EA}{{\em et al.}\xspace}
\newcommand{\TB}{\vspace{-0.1ex}}\newcommand{\TiE}{\setlength{\itemsep}{-1ex}}

\newcommand{\comment}[1]{}

\newcommand{\EG}{{\it e.g.}\xspace}
\newcommand{\FI}[1]{Fig.~\ref{#1}\xspace}

\newcommand{\BTR}{{\sf BTR}\xspace}

\newcommand{\UVX}{{u\stackrel{x}{\Rightarrow} v}}

\newcommand{\uvx}{{u\stackrel{x}{\to} v}}
\newcommand{\EC}{{\mathsf{E}_{\mathrm{fixed}}}}

\newcommand{\OO}{\mathcal{O}}
\newcommand{\HH}{\mathcal{H}}
\newcommand{\MI}{\mbox{\sf MI}}

\newcommand{\SC}{{\sf SC}\xspace}
\newcommand{\PPI}{{\sf PPI}\xspace}

\newcommand{\SCC}{{\sf\bf SCC}}

\begin{document}

\title{A New Computationally Efficient Measure of \\ Topological Redundancy of Biological and Social Networks}

\author{R\'{e}ka Albert}
    \email{ralbert@phys.psu.edu}
    \homepage{www.phys.psu.edu/~ralbert} 
    \affiliation{Department of Physics, Pennsylvania State University, University Park, PA 16802}
\author{Bhaskar DasGupta}
    \email{dasgupta@cs.uic.edu}
    \homepage{www.cs.uic.edu/~dasgupta}
    \thanks{Author to whom correspondence should be sent.}
    \affiliation{Department of Computer Science, University of Illinois at Chicago, Chicago, IL 60607}
\author{Anthony Gitter}
    \email{agitter@cs.cmu.edu}
    \homepage{www.cs.cmu.edu/~agitter}
    \affiliation{Computer Science Department, Carnegie Mellon University, Pittsburgh, PA} 
\author{Gamze G\"{u}rsoy}
    \email{gamze.gursoy@gmail.com}
    \homepage{www2.uic.edu/~ggurso2}
    \affiliation{Department of Bioengineering, University of Illinois at Chicago, Chicago, IL 60607} 
\author{Rashmi Hegde} 
    \email{rashmihegde.g@gmail.com}
    \affiliation{Department of Computer Science, University of Illinois at Chicago, Chicago, IL 60607}
\author{Pradyut Paul}
    \email{paulpradyut@yahoo.com}
    \affiliation{Junior, Neuqua Valley High School, Naperville, IL 60564}
\author{Gowri Sangeetha Sivanathan} 
    \email{gsivan2@uic.edu}
    \affiliation{Department of Computer Science, University of Illinois at Chicago, Chicago, IL 60607}
\author{Eduardo Sontag}
    \email{sontag@math.rutgers.edu}
    \homepage{www.math.rutgers.edu/~sontag}
    \affiliation{Department of Mathematics, Rutgers University, New Brunswick, NJ 08903}

\date{\today}


\begin{abstract}
It is well-known that biological and social interaction networks have a varying degree of redundancy,
though a consensus of the precise cause of this is so far lacking. In this paper, we introduce a topological 
redundancy measure for labeled directed networks that is formal,  
computationally efficient and applicable to a variety of directed
networks such as cellular signaling, metabolic and social interaction networks. We demonstrate the computational efficiency of our measure 
by computing its value and statistical significance on a number of biological and social networks with up to 
several thousands of nodes and edges.
Our results suggest a number of interesting observations:
{\bf (1)} social networks are more redundant that their biological counterparts, 
{\bf (2)} transcriptional networks are less redundant than signaling networks, 
{\bf (3)} the topological redundancy of the {\em C. elegans} metabolic network is largely due to its
inclusion of currency metabolites, and 
{\bf (4)} the redundancy of signaling networks is highly (negatively) correlated with the monotonicity 
of their dynamics.
\end{abstract}

\maketitle

\section{Introduction}

The concepts of degeneracy and redundancy are well known in information theory.
Loosely speaking, {\em degeneracy} refers to structurally different elements 
performing the same function, whereas {\em redundancy} refers to identical elements performing
the same function\footnote{We remind the reader that the term ``redundancy'' is {\em also} used in other contexts in biology 
unrelated to the definition of redundancy in this paper. For example, some researchers 
use redundancy to refer to {\em paralogous genes} that can provide {\em functional backup} for one another~\cite{KBP05}.
In addition, some researchers use the two terms, redundancy and degeneracy, interchangeably or use other terminologies for 
these concepts.}. 
In electronic systems, such measures are useful in analyzing properties such as fault-tolerance. 
It is an accepted fact that biological networks 
do {\em not} necessarily have the lowest possible degeneracy or redundancy; 
for example, the connectivity of neurons in brains suggest a high degree of
degeneracy~\cite{KW96}. 
However, as Tononi, Sporns and Edelman observed in their paper~\cite{TSE99}: 

\begin{quote}
{\sf Although many similar examples exist in all fields and levels
of biology, a specific notion of degeneracy has yet to be firmly
incorporated into biological thinking, largely because of the
lack of a formal theoretical framework}. 
\end{quote}

\noindent
The same comment holds true about redundancy as well. A further reason for the lack of incorporation of these notions in biological thinking is 
the lack of {\em effective} algorithmic procedures
for computing these measures for large-scale networks even when formal definitions are available.
Therefore, such studies are often done in a somewhat ad-hoc fashion, as in~\cite{PP04}. 
There do exist notions of ``redundancy'' in the field of analysis of 
{\em undirected} networks based on clustering coefficients (\EG, see~\cite{clus10}) or betweenness 
centrality measures (\EG, see~\cite{betw06}).
However, such notions are not appropriate for the analysis of 
biological networks where one must distinguish positive from negative regulatory interactions,  
and where the study of dynamics is of interest.

\section{Brief Review of an Information-theoretic degeneracy and redundancy measures}

Formal information-theoretic definitions of degeneracy and redundancy for dynamic biological systems 
were proposed in~\cite{TSE99} (see also~\cite{TSE94,TSE96}) based on {\em mutual-information contents}.
These definitions assume access to suitable perturbation experiments and corresponding accurate 
measurements of the relevant parameters. Thus, they are {\em not} directly comparable to the topology-based 
redundancy measures that we propose in this paper. 
Nonetheless, we next briefly review these definitions as a way to illustrate 
some key points of other measures often used in the literature that motivated us 
to define our new redundancy measure.

The authors of~\cite{TSE99} consider system consisting of $n$ elements that
produces a set of outputs $\OO$ via a fixed connectivity matrix from a subset
of these elements.
The elements are described by a jointly distributed random vector $X$ that
represents steady-state activities of the components of their system.
The degeneracy $\cD(X\,;\,\OO)$ of the system is then expressed as the average
mutual information (\MI) shared between $\OO$ and the ``perturbed''  
bi-partitions of $X$ summed over all bipartition sizes (Equation~{\bf [2b]}
of~\cite{TSE99}), \IE,
\begin{multline}\label{eq1}
\cD(X\,;\,\OO)=\frac{1}{2}\times\sum_{k=1}^n\sum_j\left(\MI^P(X_j^k\,;\,\OO)
\right. \\
\left. +\MI^P(X\setminus X_j^k\,;\,\OO)-\MI^P(X\,;\,\OO)\right)
\end{multline}
where $X_j^k$ is a $j\tx$ subset of $X$ composed of $k$ elements and 
the notation $\MI^P(\mathcal{A}\,;\,\OO)$ denotes the mutual information
between a subset of elements $\mathcal{A}$ and 
an output set $\OO$,
when $\mathcal{A}$ is injected with a small fixed amount of uncorrelated 
noise\footnote{$\MI^P(\mathcal{A}\,;\,\OO)=\HH(\mathcal{A})+\HH(\OO)-\HH(\mathcal{A},\OO)$,
where 
$\HH(\mathcal{A})$ and $\HH(\OO)$ are the entropies of $\mathcal{A}$ and $\OO$
considered 
independently, and 
$\HH(\mathcal{A},\OO)$ is the joint entropy of the subset of elements
$\mathcal{A}$ and the output set $\OO$.}; 
see~\cite{TSE99,TSE94} for details.
One can immediately see a computational difficulty in applying such a
definition: 
{\em the number of possible bipartitions could be astronomically large even
for a modest size network}.
For example, for a network with $100$ nodes which is a number smaller than all
but one of the networks considered in this paper,
the number of bi-partitions is roughly $2^{100}>10^{30}$. Measures avoiding
averaging over all
bi-partitions were also proposed in~\cite{TSE99}, but the computational
complexities and accuracies 
of these measures remain to be thoroughly investigated and evaluated on larger
networks.

In a similar manner, the redundancy \textrecipe$(X;\OO)$ of a system $X$ 
was defined in~\cite{TSE99} as the difference between summed mutual information upon perturbation between 
all subsets of size up to $1$ and $\OO$, and 
the mutual information between the entire system and $\OO$ (Equation~{\bf [3]} in~\cite{TSE99}), \IE, 
\begin{equation}\label{eq2}
\mbox{\textrecipe}(X\,;\,\OO)=\sum_{j=1}^n \MI^P(X_j^1\,;\,\OO) - \MI^P(X\,;\,\OO)
\end{equation}
Note that a clear shortcoming of this measure is that it only provides a number, but does not indicate which subset of elements are redundant.
Identifying redundant elements is important for the interpretation of results, and may also serve as an important step of 
the network construction and refinement process, as we will illustrate in our application 
to the {\em C. elegans} metabolic network and the oriented \PPI network. 
Tononi, Sporns and Edelman~\cite{TSE99} illustrated the above measure on a few model networks
as a proof of concept, but large 
networks clearly necessitate alternate measures that allow {\em efficient} calculations. 

In this paper we propose a new {\em topological} measure
of redundancy. A benefit of our new redundancy measure
is that we can {\em actually find 
an approximately minimal network} and, in the case of multiple minimal networks of similar quality, 
a subset of them by enabling a randomization step in the algorithmic procedure.
We determine this redundancy value for a number of biological and social networks of large sizes 
and observe a number of interesting properties of our redundancy measure. 

\section{Models for Directed Biological and Social Networks}

There are two very different levels of models for biological systems. 
A so-called {\em network topology} model (also known as  a ``wiring diagram''  or a ``static graph'') provides a
coarse diagram or map of the physical, chemical, or statistical
connections between molecular components of the network, 
without specifying the detailed kinetics. In this type of model, a network of
molecular interactions is viewed as a graph:  cellular components are nodes in a network, and 
the interactions between these components are represented by edges connecting the nodes. 
In this paper, we are mainly concerned with this type of model; exact details are described 
in Section~\ref{j1}.

In the other type of model, a {\em network dynamics} model,
mathematical rules (\EG, systems of Boolean rules or differential equations)
are used to specify the behavior over time of each of the
molecular components in the network. Our investigation is not directly concerned with such dynamic models.
However, since we will show a correlation of our redundancy measure for the network topology model with 
a property, namely {\em monotonicity}, of an associated network dynamics model, 
we briefly review this model in Section~\ref{j2}.

\subsection{Network Topology Model}
\label{j1}

\begin{figure}[htbp]
\epsfig{file=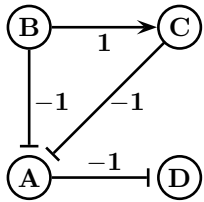}
\vspace*{-0.1in}
\caption{\label{hh}The network topology model for biological networks. The parity of the pathway 
${\text{B}\!\!\rightarrow\!\! \text{C}\!\!\rightarrow\!\! \text{A}\!\!\dashv \!\!\text{D}}$ is 
$1\!\times \!(-1)\times\! (-1)\!=\!1$.}
\end{figure}

Three common types of molecular biological networks are: {\em transcriptional regulatory} networks,
{\em metabolic} networks, and {\em signaling} networks. The nodes of transcriptional regulatory
networks represent {\em genes}, and edges represent (positive or negative) regulation of a given gene's {\em expression} 
by proteins associated to other genes. The nodes of metabolic networks are
metabolites and the edges represent the {\em enzyme-catalyzed} reactions in which these metabolites participate as reactants
or products. 
The nodes of signaling networks
are proteins and small molecules, and the edges represent physical or chemical interactions or indirect positive or negative causal effects. 
A unified formalism to describe all these types of networks uses a {\em directed} graph $G=(V,E,w)$ with vertex set $V$, edge set $E$, and 
an edge labeling function $w\,\colon\,E\mapsto \{-1,+1\}$ in which a label of $1$ (respectively, $-1$) 
represents an positive (respectively, 
negative) influence.
A pathway is then a path $P$ from vertex $u$ to vertex $v$, and the excitory or inhibitory nature of
the pathway is specified by the {\em parity} $\Pi_{e\in P}\, w(e)\in\{-1,+1\}$ of such a path $P$;
see~\FI{hh} for an illustration.

Our model for directed social interaction networks is simply a directed graph in which edges represent significant 
relationships between the entities, \EG, nodes may represent web-pages and directed edges may represent 
hyper-links of one web-page in another. Obviously, we can think of such a 
model as one of the above type in which all edges are labeled $+1$ (and, thus 
all paths have the same parity); this allows us to treat both social and biological networks 
in a mathematically uniform manner for the purpose of designing and analyzing algorithms.

\subsection{Network Dynamics and Monotonicity}
\label{j2}

Consider systems modeled via ordinary differential equations: 
\begin{equation}
\!\!\frac{\mathsf{d} x_i(t)}{\mathsf{d} t}\!=\!f_i\big(x_1(t),x_2(t),\dots,x_n(t)\big)\,\,\, \mbox{ for }i\!\!=\!\!1,2,\dots,n 
\label{system}
\end{equation}
where $x_i(t)$ indicates the concentration of the $i\tx$ entity in the model at time $t$ and the $f_i$'s are 
functions of $n$ variables.
We assume that $x(t)=(x_1(t),x_2(t),\dots,x_n(t))$ evolves in an open subset of $\R^n$, the $f_i$'s are differentiable, and 
solutions are defined for $t\geq 0$. 
For example, a simple two species interaction could be described by 
\[
\begin{array}{cl}
\frac{\mathsf{d} x_1}{\mathsf{d} t}(t) & =3 x_1(t)-5x_2(t) \\
\\
\frac{\mathsf{d} x_2}{\mathsf{d} t}(t) & =x_1(t)+x_2(t).
\end{array}
\]
A particularly appealing class of dynamics is that of 
{\em monotone} systems~\cite{Hirsch,Smithmonotone}.
Informally, the dynamics of a monotone system preserves a specific partial order (hierarchy)
of its inputs over time. 
Mathematically, monotonicity can be defined as follows.

\begin{definition}{\rm~\cite{Hirsch,Smithmonotone}}
Given a partial order $\preceq$ over $\R^n$,
system~\eqref{system} is said to be {\em monotone with respect to $\preceq$} if
\begin{multline*}
\forall\,t\geq 0 \colon\, \Big(x_1(0),\dots,x_n(0)\Big)\,\preceq\, \Big(y_1(0),\dots,y_n(0)\Big) \\
\,\,\,\,\,{\Longrightarrow} \,\Big(x_1(t),\dots,x_n(t)\Big)\,\preceq\, \Big(y_1(t),\dots,y_n(t)\Big)
\end{multline*}
where $(x_1(t),\dots,x_n(t))$ and $(y_1(t),\dots,y_n(t))$
are the solutions of~\eqref{system} with initial conditions $(x_1(0),\dots,x_n(0))$ and $(y_1(0),\dots,y_n(0))$,
respectively. 
\end{definition}

\noindent
We will restrict our attention to {\em orthant} orders.
These are the partial orders $\preceq_s$ over $\R^n$, for any given $s=(s_1,\dots s_n)\in\{-1,1\}^n$, 
defined as (see~\cite{DESZ07,Smithmonotone,Sontag:mono}): 
\[
x\preceq_s y\, {\Longleftrightarrow}\, \forall\,i\colon s_i\,x_i\leq s_i\,y_i
\]
In particular, the ``cooperative order'' is the partial order $\preceq_s$ for $s=(1,1,\dots,1)$. 

Monotone systems constitute a nicely behaved class of dynamical systems 
in several ways. For example, for these systems 
pathological behaviors (chaotic attractors) are ruled out.
Even though they may have an arbitrarily large dimensionality,
monotone systems (under an additional irreducibility assumption) behave in many ways like one-dimensional systems;
for example, bounded trajectories generically converge
to steady states, and stable oscillatory behaviors do not exist.
Monotonicity with respect to orthant orders is equivalent to the non-existence of negative 
loops in systems;  analyzing the behaviors of such loops
is a long-standing topic in biology in the context of regulation, metabolism and
development, starting from the work of
Monod and Jacob in $1961$~\cite{monod}.
In this paper, we will define a measure of ``degree of monotonicity'' for dynamical systems and relate
it to our topology-based redundancy measure.

\section{A New Measure of Redundancy} 

We will use the following notations for conciseness: 
\begin{itemize}
\item
For any two vertices $u$ and $v$, $\UVX$ (respectively, $\uvx$)
denotes a {\em path} (respectively, an {\em edge}) from $u$ to $v$ of parity $x$.
We include the empty path $u\stackrel{1}{\rightarrow} u$ for each vertex $u$. 

\item
For any $E'\subseteq E$, {\sf reachable}$(E')$ is the set of all {\em ordered}
triples $(u,v,x)$ such that $\UVX$ exists in the subgraph $(V,E')$. 
\end{itemize}
For example, for the network in \FI{hh}, 
$\text{B}\stackrel{1}{\Rightarrow}\text{D}$ exists because of the path 
$\text{B}\dashv\text{A}\dashv\text{D}$ and also because of the path
$\text{B}\rightarrow\text{C}\dashv\text{A}\dashv\text{D}$, 
and 
{\sf reachable}$\,\big(\left\{\text{B}\rightarrow\text{C},\text{A}\dashv\text{D}\right\}\big)=
\big\{
(\text{A},\text{A},1), 
(\text{B},\text{B},1), 
(\text{C},\text{C},1), 
(\text{D},\text{D},1),
\linebreak[1]
(\text{B},\text{C},1),
(\text{A},\text{D},-1), 
\big\}$.

We next state a combinatorial optimization problem that will be needed in order to introduce our new redundancy measure.

\begin{quote}
\begin{description}
\item[Problem name:]
Binary Transitive Reduction (\BTR).

\item[Instance:]
a directed graph $G=(V,E)$ with a subset of edges $\EC\subset E$ and 
an edge labeling function $w:E\mapsto \{-1,1\}$.

\item[Valid Solution:]
a subgraph 
$G'=(V,E')$ such that 
\begin{itemize}
\item
$E'\supseteq \EC$ and 

\item
{\sf reachable}$(E')=${\sf reachable}$(E)$.
\end{itemize}
($E\setminus E'$ is referred to as a set of ``redundant'' edges.)

\item[Goal:]
{\tt minimize} $|E'|$.
\end{description}
\end{quote}

\begin{figure}[htbp]
\epsfig{file=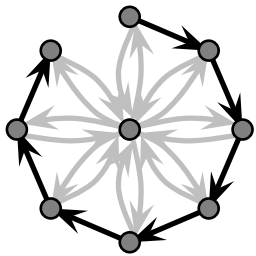}
\caption{\label{hhh}Choosing one wrong edge may cost too much in \BTR.}
\end{figure}

Intuitively, the \BTR problem prunes pathways for which alternate equivalent 
pathways exist (\EG, see~\cite{ADDKSZW07,KZSAD08}). The set of edges in $\EC$ in the definition of 
\BTR represents edges that may {\em not} be removed during the algorithm; this is 
useful in the context when one wishes to reduce a network while preserving specific pathways. 
For the redundancy calculations performed in this paper,
we assume no prior knowledge of direct interactions; thus for the rest of the paper we set $\EC=\emptyset$.
As an illustration, in \FI{hh} if we let $E'=E\setminus \{\text{B}\!\!\dashv \!\!\text{A}\}$ then {\sf reachable}$(E')=${\sf reachable}$(E)$ 
because of the path $\text{B}\!\!\rightarrow \!\!\text{C}\!\!\dashv \!\!\text{A}\!\!$. 

Finding a maximum set of edges that can be removed is non-trivial; in fact, the problem is $\NP$-hard~\cite{KRY94}. 
To illustrate the algorithmic difficulties, consider the network shown in \FI{hhh}.
Removal of all the black edges provides a non-optimal solution of \BTR, whereas an optimal solution with about 
half the edges compared to the non-optimal solution can be obtained by keeping all the black edges and removing all 
but two of the gray edges.
The special case of \BTR with $\EC=\emptyset$ and 
$w(e)=1$ for all edges $e$ is the so-called classical 
{\em minimum equivalent digraph}
problem, and it has been investigated extensively in the context of checking minimality of 
connectivity requirements in computer networks (\EG, see~\cite{KRY94}). 
Other examples of applications of \BTR-type network optimizations include 
the work by Wagner~\cite{W02} employing a special case of \BTR 
to determine network structure from gene perturbation data in the context of biological networks 
and the work by Dubois and C\'{e}cile~\cite{DB05} in the context of 
social network analysis and visualization. 

Based on the \BTR\ problem, we propose a new {\em combinatorial} measure
of redundancy that can be computed efficiently.
Note that \BTR does not change pathway level information of the network and 
removes edges from one node to another only when a
similar alternate pathway exists, thus truly removing redundant connections.
Thus, $\frac{|E'|}{|E|}$
provides a measure of global compressibility of the network
and our proposed new redundancy measure \textrecipe$_{\mathrm{new}}$ 
is defined to be 
\begin{equation}
\mbox{\textrecipe}_{\mathrm{new}}=1-\frac{|E'|}{|E|}
\end{equation}
The $|E|$ term in the denominator of the above definition 
translates to a ``min-max normalization'' of the measure~\cite{HK00}, and ensures that
$0<$\textrecipe$_{\mathrm{new}}<1$. Note that 
{\em the higher the value of \textrecipe$_{\mathrm{new}}$ 
is, the more redundant the network is}.

\subsection{Properties of Our Topological Redundancy Measure and Applications of a Minimal Network} 

Any topological redundancy measure should have a desirable property: {\em the measure must not only reflect
simple connectivity properties such as degree-sequence or average degree, it must also depend on higher-order connectivity}. 
Our redundancy measure indeed has this property, since paths of {\em arbitrary} length are considered 
for removal of an edge. For a concrete example, consider two graphs shown in \FI{red-prop}; the in-degree
and out-degree sequence of each graph is 
$\underbrace{1,1,\dots,1,1}_{\frac{n}{2}+1},\underbrace{2,2,\dots,2}_{\frac{n}{2}-1}$, 
but their redundancy values are drastically different.
Similarly, higher average degree does not necessarily imply higher values of redundancy; for example, 
the network in \FI{red-prop}, when generalized on $n$ nodes, has an average degree below $2$ and a redundancy 
value of roughly $0.33$, whereas the graph $K_{\frac{n}{2}.\frac{n}{2}}$ (a completed bipartite graph with each partition 
having $n/2$ nodes and all edges directed from the left to the right partition) has an average degree of 
$n/2$ but a redundancy value of $0$.

\begin{figure}[h]
\epsfig{file=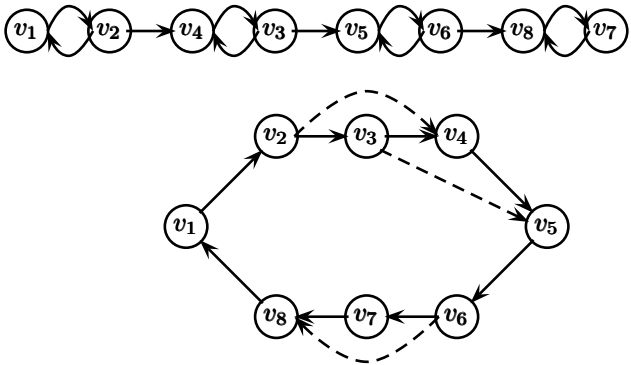}
\caption{\label{red-prop}Two $n$-node graphs with same degree sequence but with different values of \textrecipe$_{\mathrm{new}}$,
shown for $n=8$. 
The top graph has no redundant edges, thus for it 
\textrecipe$_{\mathrm{new}}=0$.
The dashed edges for the bottom graph can be removed, giving 
\textrecipe$_{\mathrm{new}}=\frac{3}{11}$.}
\end{figure}

\subsection{Computing \textrecipe$_{\mathrm{new}}$}

Although solving \BTR exactly is an $\NP$-hard problem, it has a rich combinatorial 
structure that allowed us to design an efficient approximation algorithm.
The resulting algorithms were incorporated in our {\sf NET-SYNTHESIS} software~\cite{KZSAD08} (publicly available at
\url{www.cs.uic.edu/~dasgupta/network-synthesis/}).

Although it is impossible to provide all details of the algorithmic approaches that was used 
for {\sf NET-SYNTHESIS}, we provide some high-level details of the algorithm used; the reader can find 
further details, correctness proofs and algorithmic analysis in~\cite{alb1,ADDKSZW07}.
It was proved in~\cite{alb1} 
that any strongly connected component (\SCC) of the given graph $G=(V,E)$, say $(V_1,E_1)$ with $V_1\subseteq V$ and $E_1=(V_1\times V_1)\cap E$, 
can be classified as one of the two types: 
a {\em single parity \SCC} if, for any two vertices $u,v\in V_1$,
$\UVX$ exists in the \SCC\ for exactly one $x$ from $\{-1,1\}$, and a 
{\em multiple parity \SCC} if,  for any two vertices $u,v\in V_1$,
$\UVX$ exists in the \SCC\ for both $x=1$ and $x=-1$.
A high-level view of the algorithmic approach is shown in~\FI{algm}.

The running time of {\sf NET-SYNTHESIS} is dominated by Step~{\bf 2}. Theoretically, the worst-case running 
time of the algorithm is $O(n^3)$ when $n$ is the number of vertices in $G$, but empirically 
the implementation allows us to calculate 
\textrecipe$_{\mathrm{new}}$ for networks up to about five to ten thousand nodes, thereby 
allowing us to compute the redundancy parameter for large networks.
We expect that a future 
improved implementation of \BTR will allow the calculation of redundancy 
values for even larger networks.
Regarding optimality of the computed solution, theoretically 
{\sf NET-SYNTHESIS} returns a solution that is a $3$-approximation~\cite{ADDKSZW07}, \IE
$|E_{\mathrm{solution}}|$ is no more than three times of that in an optimal solution in the worst case.
However, extensive empirical evaluations reported in~\cite{ADDKSZW07} suggest that in practice 
$|E_{\mathrm{solution}}|$ is almost always close to optimal 
(within an extra $10\%$ of the optimal). 

\begin{figure*}[htbp]
\epsfig{file=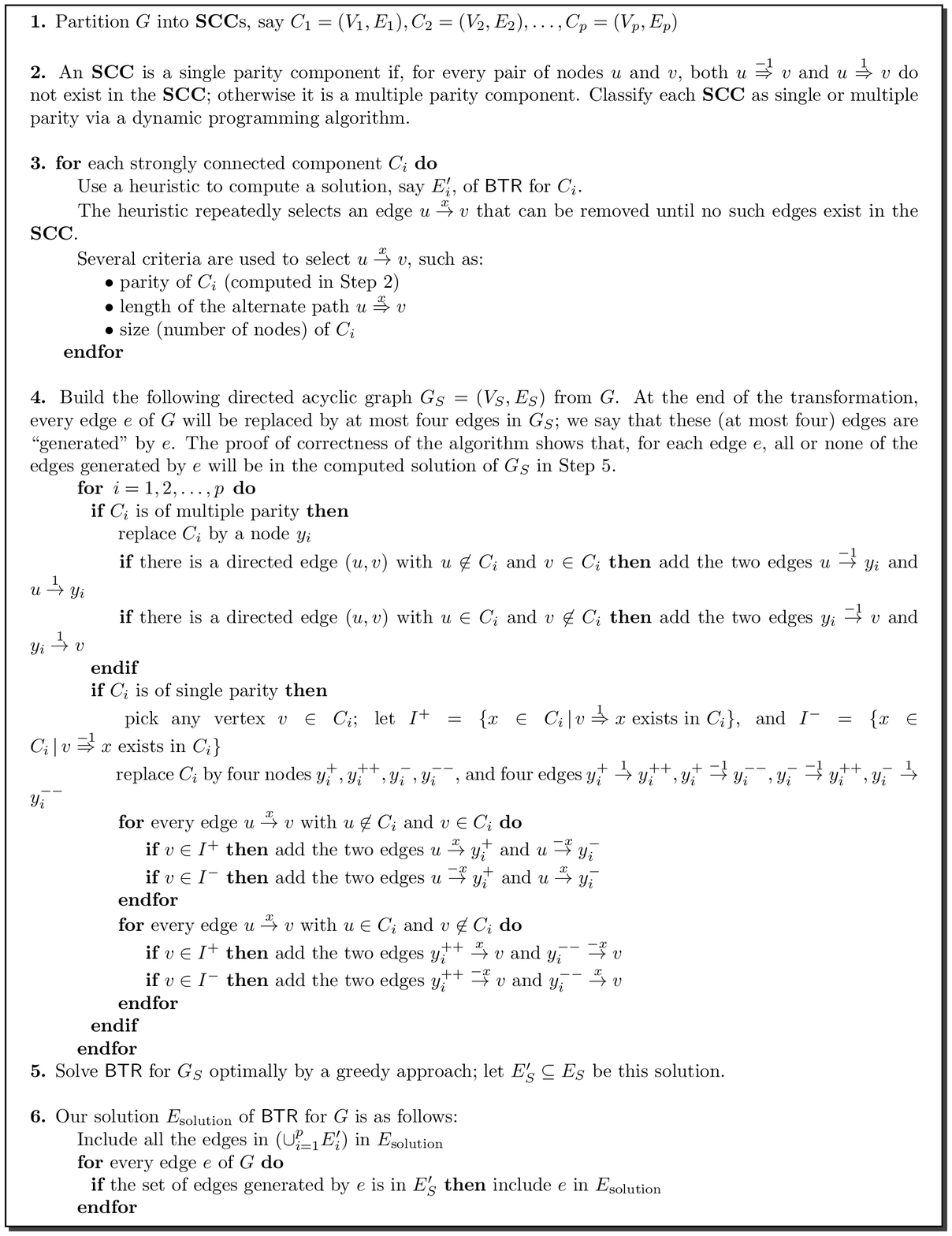}
\caption{\label{algm}A high-level view of the algorithmic approach in 
{\sf NET-SYNTHESIS} to perform \BTR.}
\end{figure*}

\begin{figure*}[htbp]
\epsfig{file=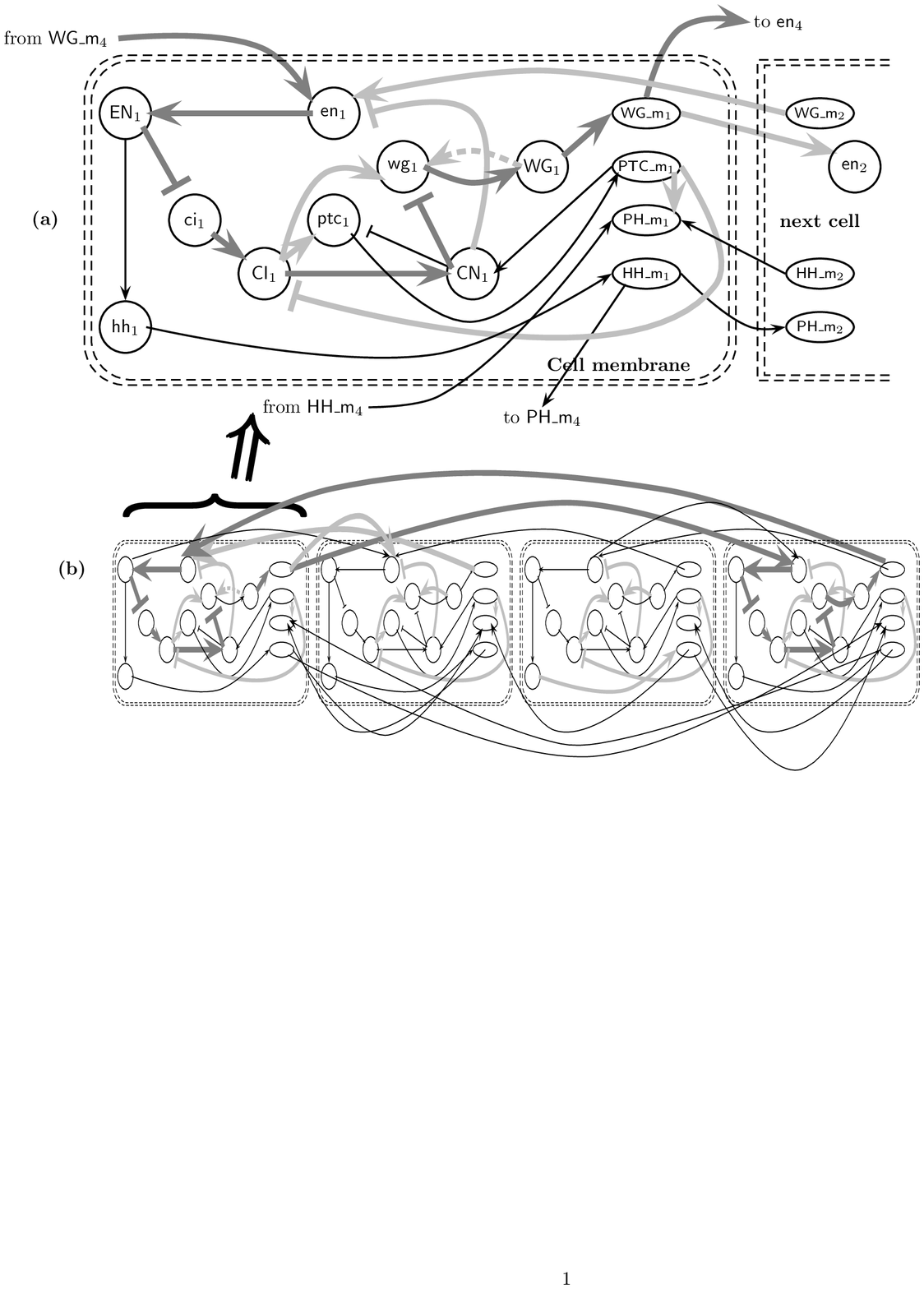}
\caption{\label{droso}{\bf (a)} The Drosophila segment polarity network for a single cell, redrawn from~\cite{Odell:2000}.
{\bf (b)} A network of $4$ cells. The redundant edges in each cell are colored light gray.
The dark gray edges form an alternate pathway of same parity for the edge {\sf WG}$_1\to\,${\sf wg}$_1$.}
\end{figure*}

\subsection{Illustration of Redundancy Calculation for a Small Biological Networks}

Our results of redundancy calculations on large-size biological and social networks are reported later, in 
Section~\ref{results}, but here we illustrate the redundancy and minimal network calculations on a biological network 
that arises from the repetition of a fixed gene regulatory network over a number of cells.
This gene regulatory network is formed among products of the segment polarity gene family, which plays an important role
in the embryonic development of {\em Drosophila melanogaster}. The interactions incorporated in this network
include translation (protein production from mRNA), transcriptional regulation, and protein-protein interactions. Two of the interactions
are inter-cellular: specifically, the proteins wingless and hedgehog can leave the cell they are produced in and
can interact with receptor proteins in the membrane of neighboring cells.
We select this network for several reasons. First, the core part of the network for a single cell is small, 
consisting of $13$ nodes and $22$ edges, which enables analytical calculations of redundancy and visual depiction 
of redundant edges. Secondly, in spite of its simplicity and regularity, the associated multi-cell network does exhibit 
non-trivial redundancies due to the inter-cellular interactions and the cyclic arrangement of cells. 
The network for a single cell was first published in~\cite{Odell:2000} and later in slightly modified form in~\cite{AO03,Ingolia04}. 
\FI{droso}{\bf (a)} shows the network of~\cite{Odell:2000} with the interpretation of the regulatory role of
{\sf PTC\_m} on the reaction {\sf CI}$\,\to\,${\sf CN} as {\sf PTC\_m}$\,\to\,${\sf CN} and {\sf PTC\_m}$\,\dashv\,${\sf CI}.
We note that the inter-cellular interactions are present at the whole cell membrane and not just the right boundary as shown for simplicity
in all reconstructions.
In a manner similar to that in other papers (\EG, see~\cite{DESZ07}), we build
a $1$-dimensional multi-cellular version by 
considering a row of $y$ cells, each of which has {\em separate} variables for
each of the compounds, letting the cell-to-cell interactions be as in \FI{droso}{\bf (a)}, but acting on both left and right
neighbors, and using cyclic
boundary conditions; see~\FI{droso}{\bf (b)} 
for an illustration.

If the network contains $y>2$ cells, then 
\begin{itemize}
\item
the number of vertices and edges are $13y$ and $22y$, respectively; and 

\item
{\sf NET-SYNTHESIS}, after performing \BTR, keeps $16y-2$ edges, giving \textrecipe$_{\mathrm{new}}=\frac{6y+2}{22y}\approx \frac{3}{11}$.
\end{itemize}
Identifying a molecule in the $i\tx$ cell via a subscript $i$, {\sf NET-SYNTHESIS} removed the following edges:
\begin{itemize}
\item
the two edges 
{\sf WG\_m}$_2\to\,${\sf en}$_1$
and 
{\sf WG\_m}$_1\to\,${\sf en}$_2$, and 

\item
the set of six edges from each cell $i\,$: 
{\sf PTC\_m}$_i \to\,${\sf PH\_m}$_i$, {\sf PTC\_m}$_i\dashv\,${\sf CI}$_i$, {\sf WG}$_i\to\,${\sf wg}$_i$, {\sf CN}$_i\dashv\,${\sf en}$_i$, 
{\sf CI}$_i\to\,${\sf wg}$_i$ and {\sf CI}$_i\to\,${\sf ptc}$_i$ 
\end{itemize}
As can be seen, the redundancies depend in a non-trivial manner on higher-order connections. For example, the light gray edge 
{\sf WG}$_1\!\to\,${\sf wg}$_1$ is redundant because of the alternate dark gray pathway shown in \FI{droso}.

\subsection{Computing the Confidence Parameter for \textrecipe$_{\mathrm{new}}$}

We apply our redundancy measure on seven biological networks
and four social networks (see Table~\ref{rr}).
For each (social or biological) network $G$ in Table~\ref{rr}, except networks {\bf (9)} and {\bf (10)}, having a redundancy value of \textrecipe$_{\mathrm{new}}(G)$, 
we generated $100$ random networks, and computed the redundancies 
\textrecipe$_{\mathrm{new}}(G_{\mathrm{random}_1})$, \textrecipe$_{\mathrm{new}}(G_{\mathrm{random}_2})$, $\dots$,\textrecipe$_{\mathrm{new}}(G_{\mathrm{random}_{100}})$
of these random networks. We then use a (unpaired) one-sample student's {\sf t}-test to determine the probability that 
\textrecipe$_{\mathrm{new}}(G)$ can be generated by a distribution that 
fits the data points \textrecipe$_{\mathrm{new}}(G_{\mathrm{random}_1})$, $\dots$, \textrecipe$_{\mathrm{new}}(G_{\mathrm{random}_{100}})$.

The current implementation of NET-SYNTHESIS runs slowly due to its intensive disk access on networks {\bf (9)} and {\bf (10)} in Table~\ref{rr} because network {\bf (9)} 
is very dense (an average degree of $9.62$ on $1133$ nodes) and network {\bf (10)} has a very large number of edges ($24316$ edges).  
Redundancy analysis of a single random graph generated for either of these two networks requires a week or more, and any meaningful statistics would require on the 
order of $100$ random graphs for each network. Due to the prohibitive time requirements we were not able to report $p$-values for these two networks
Since the characteristics of various biological and social networks are of different nature, we generate random networks for the various networks using two
different methods as explained below.

Ideally, for networks of a particular type, one would prefer to use
an accurate generative null model for highest accuracy in $p$-values.
For signaling and transcriptional biological networks (networks {\bf (1)}---{\bf (5)} in Table~\ref{rr}), 
reference~\cite{ADDKSZW07}, based on extensive literature review of similar kind
of biological networks in prior papers, arrived at the characteristics of a generative null model that is described 
below and used by us for these networks\footnote{Our simulations with the alternate Markov-chain model used for the remaining networks show that the $p$-values
still remain negligibly small; this is consistent with similar observations in another context made by Shen-Orr \EA~\cite{net1}.}. 
One of the most frequently reported topological characteristics of such networks
is the distribution of in-degrees and out-degrees of nodes, which 
exhibit a degree distribution
that is close to a power-law or a mixture of a power law and an
exponential distribution~\cite{Albert-Barabasi-2002,Giot-et-al-2003,Li-et-al-2004}. 
Specifically, transcriptional regulatory networks 
have been reported to exhibit a power-law out-degree distribution, while the in-degree distribution is more 
restricted~\cite{net1,Lee-et-al-2002}.
Based on such topological characterizations of signaling and transcriptional networks reported in the literature, 
~\cite{ADDKSZW07} used the following degree distributions for the purpose of generating random networks 
for the biological transcriptional and signaling networks such as the ones in {\bf (1)}--{\bf (5)} in Table~\ref{rr}: 
\begin{itemize}
\item
The number of vertices is the same as the network $G$ whose redundancy value was computed.

\item
The in-degree and out-degree distributions of the random networks are as follows:
\begin{itemize}
\item 
The distribution of {\sf in-degree} of the networks is {\em exponential}, \IE, 
Pr[{\sf in-degree}$=\!\!x$]$=c_1\,\mathsf{e}^{-cx}$ with 
$\frac{1}{2}<c_1<\frac{1}{3}$ 
and a maximum {\sf in-degree} of $12$. 

\item 
The distribution of {\sf out-degree} of the networks is governed by a {\em power-law}, \IE, 
for $x\geq 1$, Pr[{\sf out-degree}$=\!\!x$]$=c_2\,x^{-c}$, 
for $x=0$ Pr[{\sf out-degree}$=0$]$\geq c_2$
with $2<c_2<3$ and a maximum {\sf out-degree} of $200$.

\item
The parameters in the above distribution are adjusted such that the sum of in-degrees of
all vertices are equal to the sum of out-degrees of all vertices and the expected number of
edges is the same as $G$.
\end{itemize}
\item
The percentage for activation/inhibition edges in the random network is the same as in $G$.
\end{itemize}
Each of the $r$ random networks with these degree distributions are generated using our private implementation of  
the method suggested by Newman, Strogatz and Watts in~\cite{NSW01}. 

For social networks, for the {\em C. elegans} metabolic network and for the oriented \PPI 
network (networks {\bf (6)}--{\bf (11)} in Table~\ref{rr}), 
in the absence of a consensus on an accurate generative null model,
we generated the $r$ random networks using a Markov-chain algorithm~\cite{KTV99} in a similar manner as in, say~\cite{net1}, 
by starting with the real network $G$ and repeatedly swapping randomly chosen pairs of connections
in the following manner\footnote{Shen-Orr \EA~\cite{net1} considers swapping about $25$\% of the edges.}:

{\sf 
\begin{tabbing}
1234\=1234\=1234\=1234\=\kill
{\bf repeat} \\
\> choose two edges of $G=(V,E)$, $a\overset{x}{\to} b$ and $c\overset{y}{\to} d$, \\
\> \> \> randomly and uniformly ($x,y\in\{-1,1\}$) \\
\> {\bf if} \> $x\neq y$ or $a=c$ or $b=d$ \\ 
\> \> or $a\overset{x}{\to} d\in E$ or $c\overset{y}{\to} b\in E$ \\
\> {\bf then} discard this pair of edges \\
\> {\bf else} the random network contains the edges \\ 
\> \> $a\overset{x}{\to} d$ and $c\overset{y}{\to} b$ instead of $a\overset{x}{\to} b$ and $c\overset{y}{\to} d$ \\
{\bf until} $20\%$ of edges of $G$ has been swapped \\
\end{tabbing}
}

\section{Measure of Monotonicity for Biological Networks}

To explain the intuition behind the computation of a monotonicity measure of the dynamics
of a biological system, we start by relating the time-dynamics of the system 
with the graph-theoretic model of the network in the following way~\cite{DESZ07,Smithmonotone,Sontag:mono}.
The time-varying system as defined by Equation~\eqref{system} 
defines a labeled-graph model $G=(V,E,w)$ of the biological network 
in the following manner:
\begin{itemize}
\item 
$V=\{x_1,\ldots,x_n\}$;

\item
if $\frac{\partial f_j}{\partial x_i}\geq 0$ for all $x(t)=(x_1(t),x_2(t),\dots,x_n(t))$ and $\frac{\partial f_j}{\partial x_i}>0$ for some $x(t)$, \\
\hspace*{0.5in}then $(x_i,x_j)\in E$ and $w(x_i,x_j)=1$; 

\item
if $\frac{\partial f_j}{\partial x_i}\leq 0$ for all $x(t)$ and $\frac{\partial f_j}{\partial x_i}<0$ for some $x(t)$, \\
\hspace*{0.5in}then $(x_i,x_j)\in E$ and $w(x_i,x_j)=-1$.
\end{itemize}
(we assume that, for each $i$ and $j$, either
$\frac{\partial f_j}{\partial x_i}\geq 0$ for all $x$ 
or $\frac{\partial f_j}{\partial x_i}\leq 0$ for all $x$.) 

\begin{figure}[htbp]
\epsfig{file=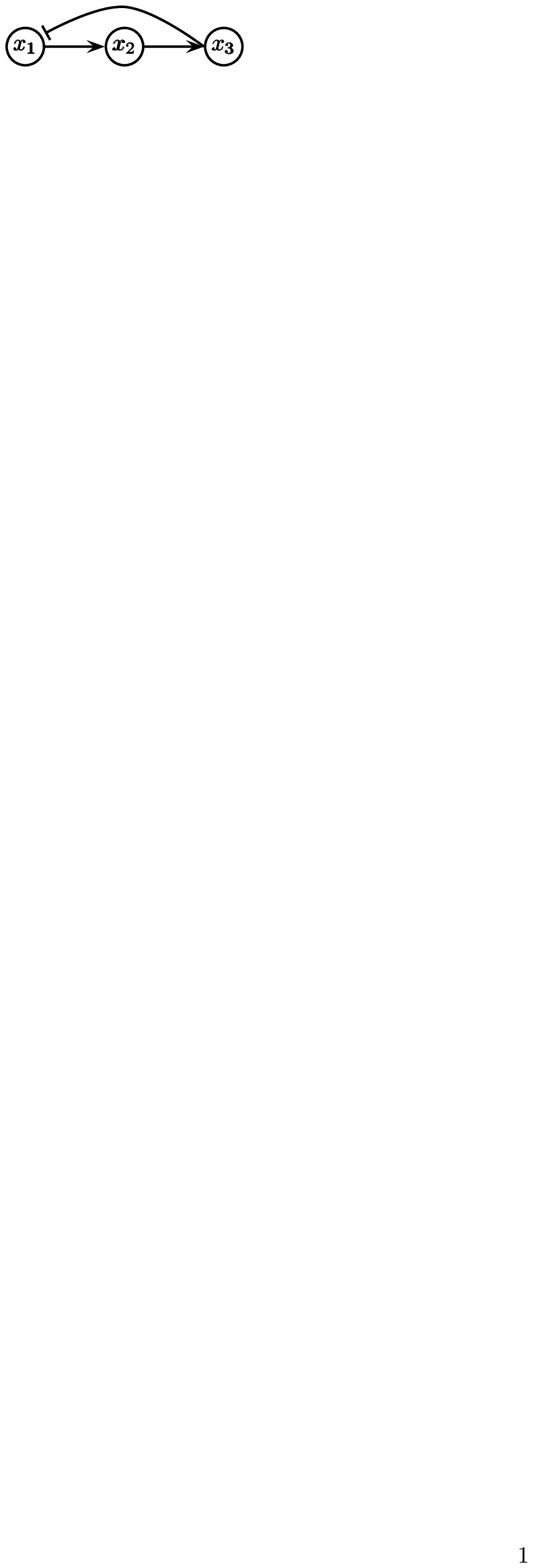}
\caption{\label{exam}\small Network for the system in Equation~\eqref{testosterone}.}
\end{figure}

As an example, consider the following biological model of testosterone
dynamics~\cite{Murray2002,Enciso:Sontag:JMB2004}:
\begin{equation}\label{testosterone}
\begin{array}{l}
\displaystyle \frac{\mathsf{d} x_1}{\mathsf{d} t}(t) =\frac{A}{K+x_3(t)}-b_1 x_1(t) \\
\\
\displaystyle \frac{\mathsf{d} x_2}{\mathsf{d} t}(t)=c_1 x_1(t) - b_2 x_2(t) \\
\\
\displaystyle \frac{\mathsf{d} x_3}{\mathsf{d} t}(t)=c_2 x_2(t) - b_3 x_3(t) 
\end{array}
\end{equation}

The corresponding labeled network for this system is shown in \FI{exam}.
It is easy to show that \eqref{testosterone} is {\em not} monotone
with respect to $\preceq_s$, for all possible $s$. 
On the other hand, if we remove the term involving $x_3$ in the first equation, we obtain a
system that is monotone with respect to $\preceq_s$, $s=(1,1,1)$.
A cause of non-monotonicity of the system 
is the existence of {\em sign-inconsistent} paths between two nodes in an {\em undirected} version 
of the network, \IE, the existence of both an activation and an inhibitory path between two
nodes when {\em the directions of the edges are ignored}. 
To be precise, define a closed {\em undirected chain} in the labeled graph $G$ as a sequence of vertices 
$x_{i_1},\dots,x_{i_r}$ such that $x_{i_1}=x_{i_r}$, and
such that for every $\lambda=1,\dots,r-1$ either
$(x_{i_\lambda},x_{i_{\lambda+1}})\in E$ or $(x_{i_{\lambda+1}},x_{i_\lambda})\in E$.
Then, the following result holds~\cite{DESZ07} (see also~\cite{deangelis} and~\cite[page 101]{smith_SIAM88}). 

\begin{lemma}{\rm~\cite{DESZ07}}\label{l1}
Consider a dynamical system~\eqref{system} with associated directed labeled graph $G$.
Then,~\eqref{system} is monotone with respect to some orthant order {\bf if and only if} 
all closed undirected chains of $G$ have parity $1$.
\end{lemma}

Note that the combinatorial characterization of monotonicity in Lemma~\ref{l1} 
is via the absence of {\em undirected} closed chains of parity $1$. Thus, in particular, any monotone system has
\begin{description}
\item[(a)]
{\em no} negative feedback loops, and 

\item[(b)]
{\em no} incoherent feed-forward-loops.
\end{description}
However, some systems may not be monotone {\em even if {\bf (a)} and {\bf (b)} hold}; see
\FI{counter-example} for an example.

\begin{figure}[htbp]
\epsfig{file=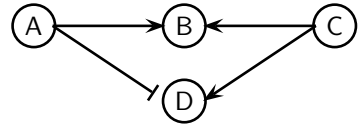}
\caption{\label{counter-example}A non-monotone system with no negative feedback loops
and no incoherent feed-forward loops.}
\end{figure}

Lemma~\ref{l1} leads in a natural manner 
to the following {\em sign consistency} (\SC) problem
to determine how monotone a system is~\cite{DESZ07,fix1}.

\begin{quote}
\begin{description}
\item[Problem name:]
Sign Consistency (\SC).

\item[Instance:]
a directed graph $G=(V,E)$ with an edge labeling function 
$w\colon E\mapsto\{-1,1\}$.

\item[Valid Solution:]
a vertex labeling function $L\colon V\rightarrow\{-1,1\}$.

\item[Goal:]
{\em maximize} $|F|$ where $F\!\!=\!\!\big\{(u,v)\,|\,w(u,v)\!\!=\!\!L(u)L(v)\big\}$
is a set of ``consistent'' edges.
\end{description}
\end{quote}

\noindent
Similar to our redundancy measure, we define the {\em degree of monotonicity} of a network to be 
\begin{equation}
{\mathsf{M}=\frac{|F|}{|E|}}
\end{equation}
where $F$ is the set of consistent edges in an optimal solution.
The $|E|$ term in the denominator of the above definition 
translates to a min-max normalization of the measure, and ensures that
$0<\mathsf{M}<1$. Note that 
{\em the higher the value of $\mathsf{M}$ is the more monotone the network is} (cf.~\cite{DESZ07,fix1}).

\subsection{Computing $\mathsf{M}$}

In~\cite{DESZ07} a semidefinite-programming (SDP) based approximation 
algorithm is described for \SC that has a worst-case theoretical guarantee of returning at least about $88\%$ of 
the maximum number of edges. The algorithm was implemented in 
{\sf MATLAB} (the {\sf MATLAB} codes are publicly available at \url{www.math.rutgers.edu/~sontag/desz_README.html}).
Other algorithmic implementations of the \SC problems are described in~\cite{fix1,fix3}.  

\subsection{Computing Correlation Between $\mathsf{M}$ and \textrecipe$_{\mathrm{new}}$}

After obtaining the ordered pair of six values 
$(\mathsf{M}_1,$\textrecipe$_{\mathrm{new}_1}),\dots,(\mathsf{M}_6,$\textrecipe$_{\mathrm{new}_6})$ 
of $\mathsf{M}$ and \textrecipe$_{\mathrm{new}}$ for the first six networks in Table~\ref{rr}, 
we computed the standard Pearson product moment correlation coefficient
$r=\dfrac{\sum_{i=1}^6 (\mbox{\textrecipe}_{\mathrm{new}_i}-\overline{\mbox{\textrecipe}_{\mathrm{new}}})(\mathsf{M}_i-\overline{\mathsf{M}})}
{\sqrt{\sum_{i=1}^6 (\mbox{\textrecipe}_{\mathrm{new}_i}-\overline{\mbox{\textrecipe}_{\mathrm{new}}})^2 \sum (\mathsf{M}_i-\overline{\mathsf{M}})^2}}$,
where $\overline{\mbox{\textrecipe}_{\mathrm{new}}}=\dfrac{\sum_{i=1}^6 \mbox{\textrecipe}_{\mathrm{new}_i}}{6}$ and 
$\overline{\mathsf{M}}=\dfrac{\sum_{i=1}^6 \mathsf{M}_i}{6}$ are the average redundancy and monotonicity values, respectively.
The possible values of $r$ always lie in the range $[-1,1]$, and  values -1 and 1 signify strongest negative and positive correlations, respectively.
A $p$-value for this correlation was calculated by a 
{\sf T}-test with two-tailed distribution and unequal variance to show 
the probability of getting a correlation as large as the observed value by random chance when the true correlation is zero.

\begin{figure}[h]
\epsfig{file=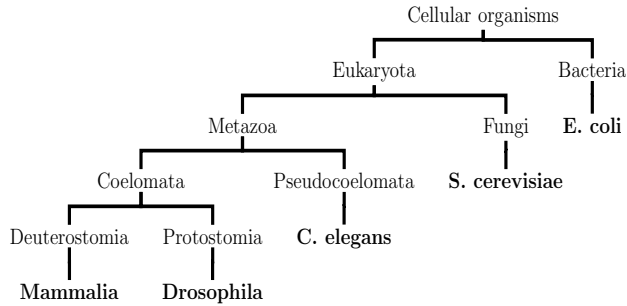}
\caption{\label{evol}An unweighted species tree of the organisms for our biological networks, 
constructed using the Taxonomy Browser resources of NCBI~\cite{ncbi}. 
The tree is not drawn to scale.} 
\end{figure}

\section{Network Data}

We selected a total of $11$ networks, seven biological ones and four social ones. We selected these networks 
with the following criteria in mind:
\begin{itemize}
\item
The biological networks were selected 
with an eye towards covering a diverse set of species on the evolutionary scale and 
towards covering networks of diverse natures (\EG, metabolic, transcriptional); a species 
tree of the biological organisms for our networks is shown in \FI{evol}.

\item
The social networks were selected covering interactions in different social environments. 

\item 
The networks span a wide range on size (number of edges ranging from $135$ to $24316$) and density (average degree ranging from $1.3$ to $13.4$) 
to demonstrate that our new redundancy measure
can be computed efficiently for a large class of networks.
\end{itemize}
Table~\ref{rr} provides more details and sources for these networks.

\begin{table*}
{\normalsize
\caption{\label{rr}Network data with sources. {\em If duplicated edges were present in the original network, they were removed in calculation of
number of edges}.}
\vspace*{0.2in} 
\begin{tabular}{m{0.3in}m{0.5in}m{0.5in}m{0.5in}m{5in}}
\hline\hline
& Number of nodes ($n$) & Number of edges ($m$) & Average degree ($m/n$) & \centerline{\raisebox{-0.2in}{\normalsize Brief Description and Reference}} \\
\hline\hline
\\
\multicolumn{5}{c}{\large\bf Biological Networks} \\
\\
\hline\hline
{\bf (1)} & $311$     & $451$     & $1.45$  &
                                    {\em E. coli} transcriptional regulatory network constructed by Shen-Orr, Milo, Mangan and Alon in~\cite{net1} for 
                                    direct regulatory interactions between transcription factors and the genes or operons they regulate; see 
                                     \url{http://www.nature.com/ng/journal/v31/n1/full/ng881.html}. \\
\hline
{\bf (2)} & $512$    & $1047$     & $2.04$  &{\em Mammalian} network of signaling pathways and cellular machines in the hippocampal CA1 neuron
                                    constructed by Ma'ayan \EA~\cite{net2}; see \url{http://www.sciencemag.org/content/309/5737/1078.abstract} \\
\hline
{\bf (3)} & $418$    & $544$      & $1.3$   & {\em E. coli} transcriptional regulatory network (updated version of the network 
                                    constructed by Shen-Orr, Milo, Mangan and Alon in~\cite{net1}); see 
                                    \url{http://www.weizmann.ac.il/mcb/UriAlon/Papers/networkMotifs/coli1_1Inter_st.txt} \\
\hline
{\bf (4)} & $59$     & $135$      & $2.28$   & {\em T cell large granular lymphocyte} (T-LGL) survival signaling network
                                    constructed by Zhang \EA~\cite{net6}; see \url{http://www.pnas.org/content/105/42/16308.abstract}.  \\
\hline
{\bf (5)} & $690$    & $1082$     & $1.56$    & {\em S. cerevisiae} transcriptional regulatory network 
                                    constructed by Milo \EA~\cite{net8} showing interactions between transcription factor proteins and genes;
                                    see \url{http://www.sciencemag.org/cgi/content/abstract/298/5594/824}.  \\
\hline
{\bf (6)} & $651$ & $2040$     &  $3.13$   & {\em C. elegans} metabolic network constructed by Jeong \EA~\cite{net10-1} and also used by
                                    Duch and Arenas in~\cite{net10-2}.  \\
\hline
{\bf (7)} & $786$ & $2453$     &  $3.12$   & An oriented version of an unweighted \PPI network constructed from {\em S. cerevisiae} interactions in the BioGRID database by 
                                 Gitter, Klein-Seetharaman, Gupta and Bar-Joseph~\cite{net11}. \\
\hline\hline
\\
\multicolumn{5}{c}{\large\bf Social Networks} \\
\\
\hline\hline
{\bf (8)} & $198$ & $2742$      &  $13.84$   & Network of Jazz musicians~\cite{net12}.  \\
\hline
{\bf (9)} & $1133$ & $10903$    &  $9.62$   & List of edges of the network of e-mail interchanges between members of the 
                                  University Rovira i Virgili (Tarragona)~\cite{email}. \\
\hline
{\bf (10)} & $11240$ & $24316$  & $2.16$  & Network of users of the Pretty-Good-Privacy algorithm for secure information interchange; edges connect users 
                                  that trust each other~\cite{net13}. \\
\hline
{\bf (11)} & $1169$ & $1912$    & $1.63$  & Enron email network; available from UC Berkeley Enron Email Analysis 
                                   (\url{http://bailando.sims.berkeley.edu/enron_email.html}). \\
\hline\hline
\end{tabular}
}
\end{table*}

\begin{table*}
{\normalsize
\caption{\label{r} 
{\bf (a)} Topological redundancy and {\bf (b)} monotonicity values. 
{\sf Higher values of \textrecipe$_{\mathrm{new}}$ (respectively, $\mathsf{M}$) imply more 
redundancy (respectively, monotonicity)}. In general, a $p$-value below $10^{-4}$ indicates statistical 
significance. N/A means not applicable; --- indicates $p$-value could not be computed
in reasonable time 
with the current implementation of {\sf NET-SYNTHESIS} because of its extensive disk access for networks that are too large or dense.
Note that the $p$-values depend not only on the average redundancies of the random networks but also on the higher order moments.}
\begin{tabular}{m{3in}||llc||c} 
\multicolumn{1}{c}{} & \multicolumn{3}{c}{\bf (a)} & {\bf (b)} \\
\hline\hline
& \multicolumn{3}{c||}{Redundancy} & Monotonicity \\
\multicolumn{1}{c||}{Network} & \textrecipe$_{\mathrm{new}}$ & \multicolumn{1}{c}{$p$-value} & \multicolumn{1}{c||}{{{{average redundancy}}}}    & $\mathsf{M}$ \\ 
        &                              &                               & \multicolumn{1}{c||}{{{{of random networks}}}}    &              \\ 
\hline\hline
\multicolumn{5}{c}{} \\ 
\multicolumn{5}{c}{\large\bf Biological Networks} \\ 
\multicolumn{5}{c}{} \\ 
\hline\hline
{\bf (1)}$\,$ {\em E. Coli} transcriptional & $\,\,0.062$  & $\,\,\,\,1.43\times 10^{-29}$ & $0.188$ & $0.796$ \\ 
\hline
{\bf (2)}$\,$ {\em Mammalian} signaling & $\,\,0.434$  & $\,\,\,\,4.4\times10^{-52}$   & $0.576$ & $0.593$ \\
\hline
{\bf (3)}$\,$ {\em E. Coli} transcriptional & $\,\,0.068$ & $\,\,\,\,2.61\times 10^{-9}$ & $0.099$ & $0.862$ \\
\hline
{\bf (4)}$\,$ T-LGL signaling       & $\,\,0.438$  & $\,\,\,\,1.15\times 10^{-11}$ & $0.350$ & $0.867$ \\
\hline
{\bf (5)}$\,$ {\em S. cerevisiae} transcriptional & $\,\,0.060$  & $\,\,\,\,9.34\times 10^{-43}$ & $0.228$ & $0.926$  \\
\hline
{\bf (6)}$\,$ {\em C. elegans} metabolic & $\,\,0.669$  &  $\,\,\,\,2.2\times 10^{-147}$ & $0.790$ & $0.444$ \\
\hline
{\bf (7)}$\,$ Oriented {\em S. cerevisiae} protein interactions & $\,\,0.481$  & $\,\,\,\,3.68\times 10^{-111}$  & $0.593$ &   N/A \\
\hline\hline
\multicolumn{5}{c}{} \\ 
\multicolumn{5}{c}{\large\bf Social Networks} \\ 
\multicolumn{5}{c}{} \\ 
\hline\hline
{\bf (8)}$\,$ Jazz musicians network & $\,\,0.897$  &  $\,\,\,\,1.06\times 10^{-107}$               &   $0.929$ & N/A \\
\hline
{\bf (9)}$\,$ Email network at University Rovira i Virgili & $\,\,0.840$  & \multicolumn{1}{c}{---}  & \multicolumn{1}{c||}{---} & N/A \\
\hline
{\bf (10)}$\,$ Secure information interchange user network & $\,\,0.486$  & \multicolumn{1}{c}{---}  & \multicolumn{1}{c||}{---} & N/A  \\
\hline
{\bf (11)}$\,$ Enron email network & $\,\,0.352$  & $\,\,\,\,2.14\times 10^{-68}$ &  $0.377$ & N/A  \\         
\hline\hline
\end{tabular}
}
\end{table*}

\section{Results and Discussions}
\label{results}

In Table~\ref{r} we show the tabulation of redundancy and, when appropriate, also monotonicity values
for our networks. Because of their large sizes, $p$-values for the redundancy measure could not be estimated
very reliably for networks {\bf (9)} and {\bf (10)}
since they require runs on many random networks, each of which would take upwards of a week; thus we do not report $p$-values for these networks. 
The extremely low $p$-values in Table~\ref{r} indicate that the real networks' redundancy values cannot be generated by a distribution that fits the 
redundancies of the equivalent random graphs.

\begin{table*}[htbp]
\caption{\label{hhhh}Normalization keeps relative magnitudes and ranks of values similar to that in the original.}
\begin{tabular}{l|l|l|l|l|l|l|l|l|l}\hline\hline
& \multicolumn{9}{c}{Networks} \\ \cline{2-10}
& {\bf (1)} & {\bf (2)} & {\bf (3)} & {\bf (4)} & {\bf (5)} & {\bf (6)} & {\bf (7)} & {\bf (8)} & {\bf (11)} \\ \hline
& & & & & & & & & \\
Original Redundancy \textrecipe$_{\mathrm{new}}$ & 0.062 & 0.434 & 0.068 & 0.438 & 0.06 & 0.669 & 0.481 & 0.897 & 0.352 \\
& & & & & & & & & \\
Normalized Redundancy $\widehat{\text{\textrecipe}_{\mathrm{new}}}$ & 0.048 & 0.364 & 0.070 & 0.319 & 0.043 & 0.708 & 0.497 & 1.112 & 0.295 \\
\hline\hline
\end{tabular}
\end{table*}

If one prefers, a normalization of the redundancy values of the networks for which randomly generated networks are available can be performed 
as follows.
For each of the nine networks, we first computed the standardized redundancy value for each of the $100$ random networks to eliminate sampling bias 
(for a sample $x_1,x_2,\dots,x_m$ with average $\mu$ and standard deviation $\sigma$, the standardized value of $x_i$ is given by $\frac{x_i-\mu}{\sigma}$).
Then, we calculated the standardized range (difference between maximum and minimum) of these $100$ standardized redundancy values. Finally, we normalized
original redundancy value by dividing them by this standardized range. The resulting 
normalized values are shown in Table~\ref{hhhh} (for comparison purposes, the normalized redundancy values are scaled so that their summation is exactly the same as the summation 
of original redundancy values). As can be seen, the ranks of both original and normalized values are almost the same (in the order 
{\bf (5)}, {\bf (1)}, {\bf (3)}, {\bf (11)}, {\bf (2)}, {\bf (4)}, {\bf (7)}, {\bf (6)}, {\bf (8)} 
and {\bf (5)}, {\bf (1)}, {\bf (3)}, {\bf (11)}, {\bf (4)}, {\bf (2)}, {\bf (7)}, {\bf (6)}, {\bf (8)}, respectively)  
and the relative magnitudes of the values are similar whether one uses
the normalized or original values, and thus all of our conclusions are valid in either case.
Thus, in the rest of the paper, we use the original redundancy values with the understanding that 
all of our conclusions are valid for the normalized values as well.

In spite of our somewhat limited set of experiments, our results do point to some interesting hypotheses, 
which we summarize below.

\subsection{\textrecipe$_{\mathrm{new}}$ can be computed quickly for large networks and is statistically significant}

As our simulations show, the new redundancy measure can be computed quickly for networks up to thousands of nodes;
for example, typically {\sf NET-SYNTHESIS} takes from a few seconds up to a minute for networks having up to $1000$ nodes or edges.
This is a desirable property of any redundancy measure so that it can be used by future researchers as 
biological and social networks grow in number and size. Moreover, the extremely low $p$-values suggests statistical significance of the new measure.

\subsection{Redundancy variations in biological networks}

We focus our attention to the variations of the redundancy values for the five transcriptional/signaling biological 
networks in our dataset and make the following observations.

\paragraph{Transcriptional vs. signaling networks}
Networks {\bf (1)}, {\bf (3)} and 
{\bf (6)} are transcriptional networks with 
all having similar low redundancies ($0.062$, $0.068$ and $0.06$).
On the other hand, 
network {\bf (2)} is a signaling
network and network {\bf (4)} is also {\em predominantly} signaling, though it
includes four transcriptional edges; these 
two mammalian signal transduction networks have similar mid-range
redundancies, namely $0.434$ and $0.438$, respectively.
We hypothesize that in general transcriptional networks are less redundant
than signaling networks. 
A straightforward supporting evidence for this
is the higher average degree of signaling networks as compared to
the transcriptional ones.
Transcriptional networks have indeed been reported to have a
feed-forward structure with few feedback loops and relatively low
cross-talk~\cite{BBO05}, whereas~\cite{net2}
reports a large
strongly connected component for their studied signaling networks (which makes it possible to reach almost
any node from any input node).

\paragraph{Role of currency metabolites in redundancy of metabolite networks}\label{met-sec}
Our data-source for the {\em C. elegans} metabolic network includes two types
of nodes, the {\em metabolites} and {\em reaction} nodes, and the edges are directed
either from those metabolites that are the reactants of a reaction to the reaction
node, or from the reaction node to the products of the reaction.
In this representation, redundant edges  appear if both (one of) the
reactant(s) and (one of) the product(s) of a reaction appear as
reactants  of a different reaction, or conversely, both (one of) the
reactant(s) and (one of) the product(s) of a reaction appear as
products  of a different reaction. Because a reaction cannot go
forward if one of its reactants is not present, the redundant edges
are not biologically redundant and cannot be eliminated. Our result of
a surprisingly high redundancy value for the metabolic network
nevertheless indicates a high abundance of a pattern, which warrants
further investigation.

One possibility we considered is that one of the reactions is
essentially a {\em dimerization} of a compound and its slightly modified
variant. However, we found no strong support for this case. Another
possibility  is that metabolites that participate in a large number of
reactions will have a higher chance to be the reactant or product of
such ``redundant'' edges. There is a biological basis for this possibility
in the existence of {\em currency metabolites}.
Currency metabolites (sometimes also referred to as {\em carrier} or {\em current}
metabolites) are plentiful in normally functioning cells and occur in
widely different exchange processes. For example, {\sf ATP} can be seen as
the energy currency of the cell. Because of their wide participation
in diverse reactions, currency metabolites tend to be the highest
degree nodes of metabolic networks. There is some discussion in the
literature on how large the group of currency metabolites is, but the consensus
list includes $\mathsf{H_20}$, {\sf ATP}, {\sf ADP}, {\sf NAD} and its variants, {\sf NH4+}, and {\sf PO4}$^{3-}$
(phosphate)~\cite{currency1,currency2}.

Our data source for the {\em C. elegans} metabolic network indicates the
identity of the $10$ highest in-degree nodes (as a group) and the $10$
highest out-degree nodes (as a group). Out of the $13$ distinct nodes in the
aggregate of these two groups, $11$ belong in the consensus list of
currency metabolites, leaving out {\sf co-enzyme A} and {\sf L-glutamate}. We found
that when we rank the nodes of the network by the number of redundant edges (as found 
by {\sf NET-SYNTHESIS}) incident upon them 
and consider the top $17$ nodes in this rank order, 
they include all the $13$ highest degree nodes in the original networks.
Thus we can conclude that the topological
redundancy of the {\em C. elegans} metabolic network is largely due to its
inclusion of currency metabolites.

\subsection{Redundancy of Social vs. Biological Networks}

\begin{figure}
\epsfig{file=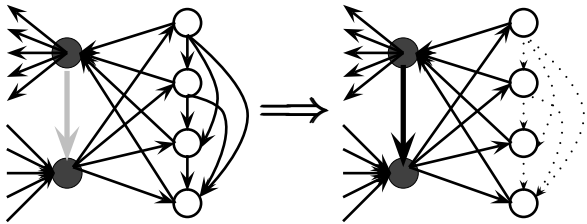}
\caption{\label{inc}Adding the edge colored light gray may increase the redundancy of the social network drastically (removed edges shown
as dotted).}
\end{figure}

The results in Table~\ref{r} seem to suggest that social networks are more redundant than biological networks.
In fact, the two most redundant networks in the table are the two social
networks {\bf (8)} and {\bf (9)} which have redundancies about twice than that of any biological networks considered,
and the remaining two social networks have redundancies comparable to the highest redundancy of the biological networks.
We hypothesize that in general this is the case. This hypothesis is perhaps not very surprising in the 
context of past research as explained below.

The research work of Navlakha and Kingsford~\cite{NK11}
suggests that biological networks may grow and evolve {\em quite differently} than social networks. 
In particular, they show that models for biological networks may perform poorly for social networks and vice versa.
It is conceivable that different models may give rise to different magnitudes of redundancy.

\begin{table*}[htbp]
\begin{center}
\caption{\label{assort}\label{tau}Values of the assortativity coefficient $r$ and the transitivity coefficient $\tau$. Negative values of $r$ 
indicate disassortativity whereas positive values of $r$ indicate assortativity.}
{\normalsize
\begin{tabular}{rrrrrrrrrrrr} \hline\hline
& \multicolumn{11}{c}{Network Index} \\ \hline
& \multicolumn{7}{c|}{Biological} & \multicolumn{4}{c}{Social} \\ \hline\hline
& {\bf (1)} & {\bf (2)} & {\bf (3)} & {\bf (4)} & {\bf (5)} & {\bf (6)} & \multicolumn{1}{c|}{{\bf (7)}} & {\bf (8)} & {\bf (9)} & {\bf (10)} & {\bf (11)} \\ \hline
$r=\,\,\,$ & $-0.149\,\,\,$ & $-0.106\,\,\,$ & $-0.204\,\,\,$ & $-0.089\,\,\,$ & $-0.398\,\,\,$ & $-0.060\,\,\,$ & \multicolumn{1}{r|}{$-0.1377\,\,\,$} & $+0.02\,\,\,$ & $+0.07\,\,\,$ & $+0.239\,\,\,$ & $-0.44\,\,\,$ \\ \hline
$\tau=\,\,\,$ & $0.037\,\,\,$ & $0.010\,\,\,$ & $0.007\,\,\,$ & $0.043\,\,\,$ & $0.005\,\,\,$ & $0.047\,\,\,$ & \multicolumn{1}{r|}{$0.017\,\,\,$} & $\mathbf{0.255}\,\,\,$ & $0.058\,\,\,$ & \multicolumn{1}{c}{---} & $0.013\,\,\,$ \\ \hline\hline
\end{tabular}
}
\end{center}
\end{table*}

Some previous research works (\EG, see~\cite{soc1,soc2,soc3}) 
ascertain that social networks tend to exhibit {\em assortativity} (\IE, highly connected nodes tend to be 
connected with other high degree nodes), whereas  
biological networks typically show {\em dissortativity} (\IE, high degree nodes tend to attach to low degree nodes). 
It is not difficult to see that such properties may lead to the difference in redundancies for the two types of networks; 
For example, in~\FI{inc} an edge between two nodes of high degree results in removal of a large number of edges.
To check the general hypothesis of assortativity for our specific networks, we computed the assortativity 
coefficient for a network as defined in~\cite{soc2}. This coefficient is calculated in the following manner.  
First, we ignore the direction of edges obtaining an undirected graph $G=(V,E)$ from the given directed graph.
Then, the assortativity coefficient $r$ is computed by the following formula:
\[
\textstyle
r = 
\frac
{\frac{1}{|E|} \sum_{\{u,v\}\in E} \deg_u \deg_v - \left[\frac{1}{2\,|E|}\sum_{\{u,v\}\in E} (\deg_u+\deg_v)\right]^2}
{\frac{1}{2\,|E|}\sum_{\{u,v\}\in E}\left[(\deg_u)^2+(\deg_v)^2\right] - \left[\frac{1}{2\,|E|}\sum_{\{u,v\}\in E} \left(\deg_u+\deg_v\right)\right]^2}
\]
where $\deg_u$ denotes the degree of a node $u$. 
It is known that $-1\leq r\leq 1$, and more negative (respectively, more positive) values of $r$ indicating 
more disassortativity (respectively, more assortativity) of the given network.
As Table~\ref{assort} shows, all biological networks are disassortative, whereas all but one social network 
are assortative.

Finally, social networks that are related to human behavior are often expected to exhibit a high degree 
of transitivity~\cite{HL98,KT08,L73}. For example, the classical work of Leinhardt~\cite{L73} asserts that the 
structure of interpersonal relations in children's groups will progress in consistent fashion from less to more transitive organization as the 
children become older. Transitivity in this type of behavioral context translates to
coherent type~$1$ feed-forward loops (\IE, feed-forward loops of the form $A\to B$, $B\to C$ and $A\to C$), each of which contains a 
redundant edge, and thus higher transitivity immediately implies higher redundancy in our context.
To check how far this general hypothesis holds for our specific networks, we calculated the transitivity coefficient 
for our networks. The transitivity coefficient $\tau$ of a directed network~\cite{WF94} is given by 
$\frac{\mu_3}{\mu_2+\mu_3}$ where $\mu_2$ and $\mu_3$ are the number of {\em ordered} triplets 
of vertices that has two and three edges among them, respectively. We used an obvious algorithm to calculate
this value; $\tau$ could not be calculated within reasonable time for the social network {\bf (10)} in Table~\ref{rr} 
because of its large number of nodes and edges. As shown in Table~\ref{tau}, 
all the biological networks have small transitivity coefficients, and among the social networks, 
network {\bf (8)} has a value of $\tau$ that is significantly more than any of the biological networks.

\subsection{Redundancy, minimality and orienting \PPI networks}

\begin{figure}
\epsfig{file=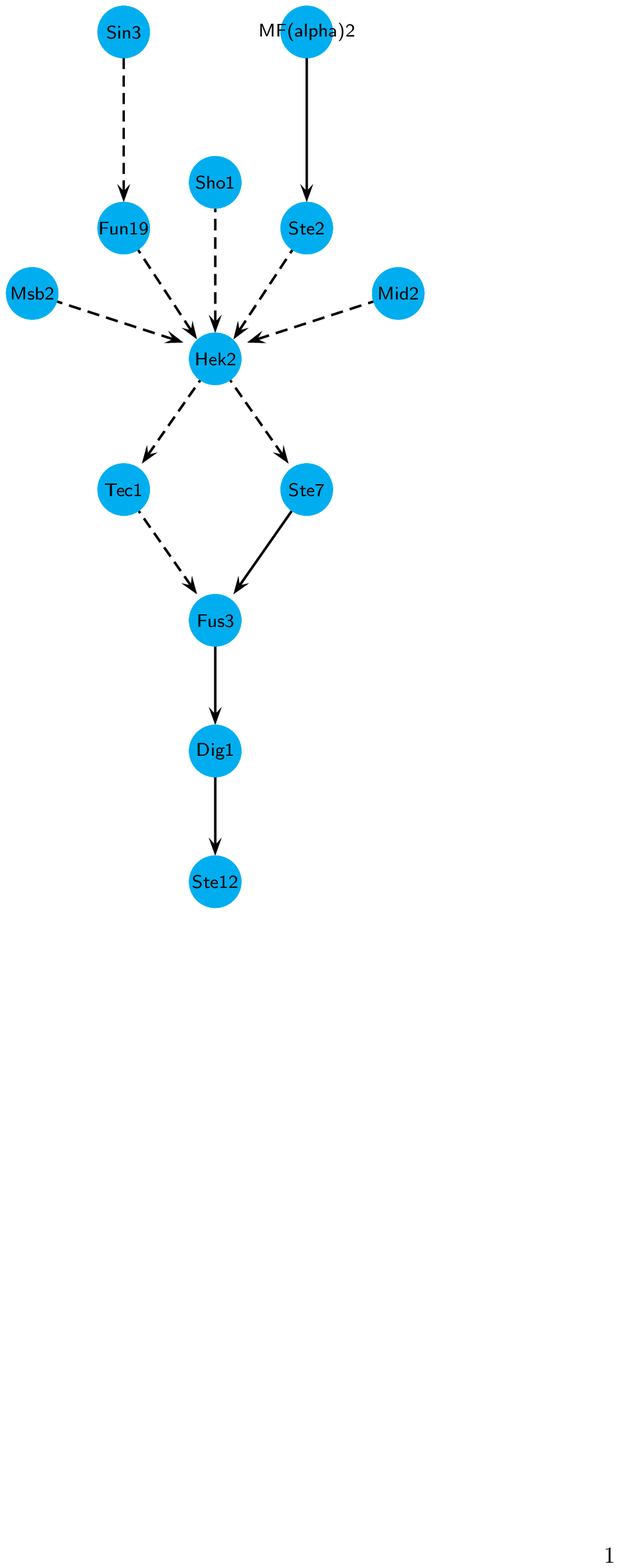}
\caption{\label{four}(color online) Paths in the non-redundant oriented \PPI network that match known yeast signaling pathways.  
Solid edges are present in the gold standard and dashed edges represent novel predictions.}
\end{figure}

Protein interaction networks represent {\em physical} interactions among proteins. 
While many protein interactions have an orientation, the current maps of protein-protein interaction (\PPI) networks 
are often unoriented (undirected) in part due to the limitations of the 
current experimental technologies such as~\cite{F05}. Thus, there is an obvious interest in trying to orient these networks by, say, combining
causal information at the cellular level. Unfortunately, most versions of the orientation problem 
is theoretically $\NP$-hard~\cite{AH02,MBZS08}, and thus heuristics for such orientations 
may either not lead to all pathways of interest or lead to extra spurious pathways that are not supported~\cite{MBZS08,net11}.

Our calculation of redundancy values and minimal networks 
provides a way to gain insight into a predicted orientation 
of a \PPI network and to determine whether the predicted oriented network has a level of redundancy similar to those in known 
biological networks.  Obviously, the lower the value of
\textrecipe$_{\mathrm{new}}$ is, the more compact is the construction of the oriented network.
However, one must also ensure that the minimal network also contains the right kind of pathways, 
\EG, paths in the ``gold standard''.  
To this effect, we describe the results of this approach via the {\sf NET-SYNTHESIS} software on an oriented \PPI 
network from~\cite{net11}.

We first briefly review the method by which the oriented \PPI network  used by us was generated.
The starting point for the network consisted of all physical interactions among yeast proteins from version 
$2.0.51$ of BioGRID~\cite{nu1}. Edge weights were assigned based on the type and quantity of experimental support for each interaction, 
and low-weight edges were removed from the network. The network was oriented so as to maximize the weighted number of 
length-bounded paths between predetermined sources and targets, which were taken from yeast MAPK signaling pathways.  
The final set of $2435$ edges included all oriented edges that belonged to any path with $5$ or fewer edges between a source and target and 
edge weights were dropped for subsequent analysis. The sources, targets, \PPI filtering and orientation algorithm are described more fully in~\cite{net11}.

Now we discuss the paths in the non-redundant network (after reduction via {\sf NET-SYNTHESIS}) that are present in the gold standard.  
Several of the short source-target paths in this network correspond to known yeast MAPK signaling pathways, specifically 
the pheromone response and filamentous growth pathways (\url{www.genome.jp/kegg/pathway/sce/sce04011.html}).  
\FI{four} depicts the union of all linear paths in the non-redundant network that have multiple consecutive edges that 
match a gold standard path.
The paths that matched a gold standard path are {\em highly similar}, and the common gold standard edges in these hits are 
{\tt Ste7$\to$Fus3}, {\tt Fus3$\to$Dig1} and {\tt Dig1$\to$Ste12}. 

\subsection{Correlation between redundancy and network dynamics}
\label{mono-sec}

\begin{figure}
\epsfig{file=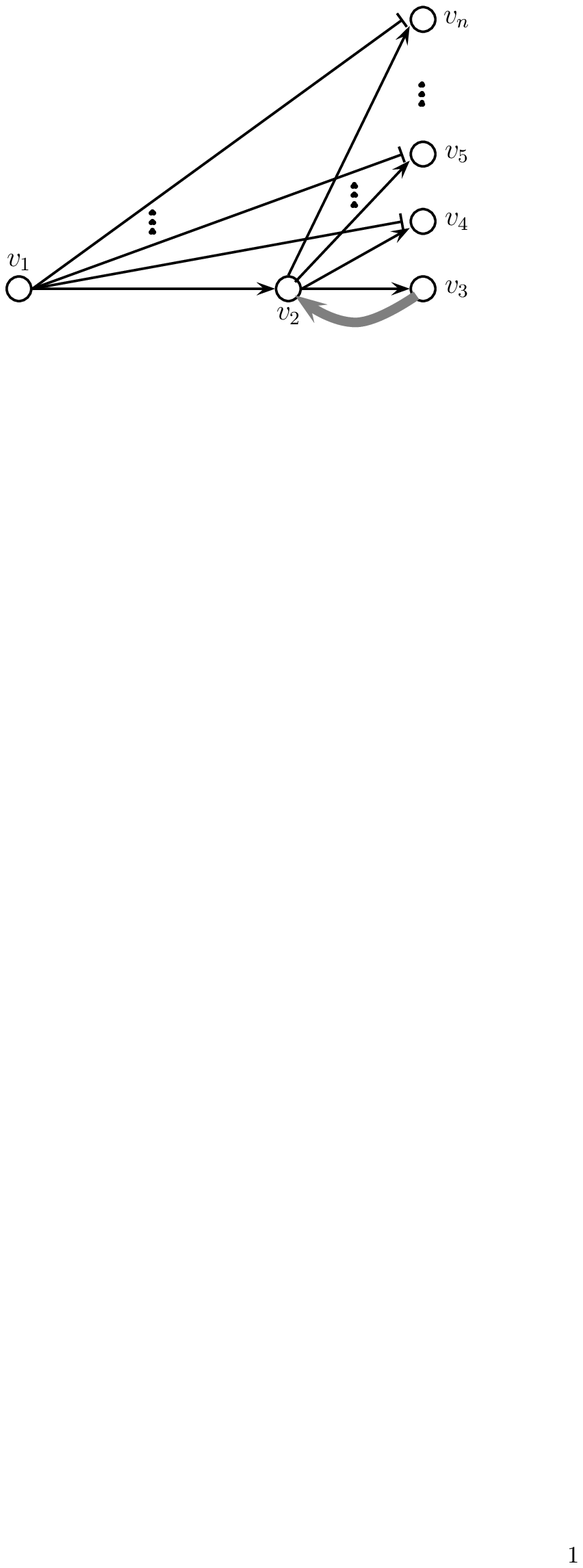}
\caption{\label{correlation}The network shown has no negative feedback loops and no redundant edges.
However, if we replace the gray activation edge $v_3\to v_2$ to an inhibition edge $v_3\dashv v_2$, a negative feedback loop
is created and this makes all the remaining inhibitory edges in the network redundant (\EG, the edge $v_1\dashv v_4$ is redundant
because of the path $v_1\to v_2\to v_3\dashv v_2\to v_4$).}
\end{figure}

The Pearson correlation coefficient between $\mathsf{M}$ and \textrecipe$_{\mathrm{new}}$ is about $-0.8$ with 
a $p$-value of $0.0066$. Thus, monotonicity is negatively correlated to redundancy (\IE, higher values 
of redundancy are expected to lead to lower values of monotonicity and vice versa). 

As explained before, monotonicity is known to be negatively correlated to 
negative feedback loops~\cite{DESZ07,halsmithJDE}. Negative feedback loops also tend 
to increase the redundancy of signal transduction networks; 
see \FI{correlation} for an illustration. Indeed, strongly connected components with at least one negative feedback loop were called a 
multiple parity components in~\cite{alb1} and played a significant role in redundancy calculations. 

Furthermore, recent results of Kwon and Cho~\cite{KC08} on the correlation
between topological properties and robustness of networks are also consistent
with the negative correlation that we obtained.  The authors of that paper
considered a weighted network model in which the state of each node is a real
number in the range $\{-1,1\}$ and the positive and negative weights of the
connections represent the strengths of the excitory or inhibitory connections,
respectively.  A negative (respectively, positive) feedback loop is then
defined to be a simple cycle with odd (respectively, even) number of negative
weights in the cycle, and the degree of robustness of a network is then
defined by selecting a group of nodes randomly, perturbing the values of their
states, and measuring the extent of change of states of various nodes in the
network by computing the ratio of state values converging to a same final
state to which the original initial state converged (biologically, this
concept of robustness means the extent of maintaining the original stable
state against given perturbations).  Based on extensive simulation results,
the authors concluded that networks with fewer negative feedback loops are
likely to be more robust in their sense. More robustness with respect to
perturbations suggests less influence of one node on another, and consequently
fewer alternate pathways of the {\em same nature} from a node to another,
indicating less redundancy values, whereas fewer negative feedback loops
correspond to higher degree of monotonicity. Thus, their observation is, at
least on an intuitive level, consistent with our finding.

\subsection{Significance of a minimal network}

It is certainly an interesting question to ask if a topologically minimal network has similar dynamical or functional properties as the 
original network.  Note that the question does not make sense for the four (static) social networks (networks {\bf (8)}, {\bf (9)}, {\bf (10)} and {\bf (11)} in Table~\ref{rr}), 
since the individual nodes in these networks usually do not have well-defined functions or dynamics, and 
one of their {\em most interesting} properties, namely connectivity, {\em is preserved} in the minimal network. 
The redundancy issue of the metabolic network (network {\bf (6)} of Table~\ref{rr}) is explained separately in detail 
in Section~\ref{met-sec}. There is no associated dynamics with the oriented \PPI network (network {\bf (7)} of Table~\ref{rr}).
Thus, this question {\em only applies} for the first five biological networks (networks {\bf (1)}, {\bf (2)}, {\bf (3)}, {\bf (4)} and {\bf (5)}) in Table~\ref{rr}.
A dynamic description/model of these networks would characterize dynamic behaviors, such as stability and 
response to external inputs. When the network has designated outputs or read-outs, such as gene expression rates in 
transcriptional networks, it may be of interest to characterize the behavior of these outputs as a function of the inputs.

A topologically minimal network has the same input-output connectivity (reachability) as the original and thus the excitory or
inhibitory influence between each input-output pair {\em is} preserved.  
It is minimal in the ``information theoretic'' sense in that any network with the same output behavior must be 
of at least this size.
A correlation of the redundancy measure with the monotonicity of dynamics is explored
in Section~\ref{mono-sec}. 
Will a topologically minimal network also have the same output behavior as the original one for the same input?  
In general, there is no such guarantee since the dynamics depend on what type of functions (``gate'') are used 
to combine incoming connections to nodes and the ``time delay'' in the signal propagation, both of which
are omitted in the graph-theoretic representation of regulatory and signal-transduction networks such as {\bf (1)--(5)} in Table~I.
For example, consider the two networks shown in \FI{figtwographs} in which 
network~{\bf (b)} has a redundant connection $A\rightarrow C$. The functions of these two circuits could be different, however, 
depending on the ``gate'' function used to combine the inputs $B\rightarrow C$ and $A\rightarrow C$ in network~{\bf (b)}.
Due to the shared $A\rightarrow B\rightarrow C$ connectivity in the two networks, in both cases node $C$ will be activated if $A$ is {\em continuously supplied}.  
However, while network~{\bf (a)} merely implements a delay between $C$ and $A$,  the coherent type-1 feedforward loop indicated in {\bf (b)} is what~\cite{alonsbook} 
calls a ``sign-sensitive delay element'' that filters spikes in signals (low-pass filter) {\em provided} that an 
``AND'' gate combines the inputs to node $C$; one example of such a circuit is that of the Arabinose system in 
{\em E.coli}~\cite{mangan2003}. In summary, deleting edges may result in functionalities that are not exactly the same.

\begin{figure}[htbp]
\epsfig{file=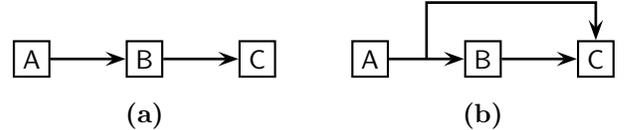}
\caption{\label{figtwographs}Equivalence of dynamics depends on node functions.}
\end{figure}

However, despite the fact that a minimal network may not preserve {\em all} dynamic properties of the original one, 
a significant application of finding minimal networks lies precisely in allowing one to identify redundant connections (edges).  
In this manner, one may focus on investigating the functionalities of these redundant edges, \EG, identifying the manner in which their effect is 
cumulated with those of the other regulators of their target nodes could be a key step toward understanding the behavior of the entire network.

Thus, the tools developed here are of general interest as they not only provide a quantified measure 
of overall redundancy of the network, but also also allow their identification of redundancies and hence help direct future research toward the 
understanding of the functional significance of the added links.

\section{Availability of Data and Software}

Most of the data for the original network as well as those for the random networks used in the calculation of $p$-values for 
\textrecipe$_{\mathrm{new}}$ are available from our website 
\url{www.cs.uic.edu/~dasgupta/network-data/}.
The {\sf NET-SYNTHESIS} software for calculating redundancies is available from our website 
\url{www.cs.uic.edu/~dasgupta/network-synthesis/}.
MATLAB codes for computing monotonicity values
are available from our website \url{www.math.rutgers.edu/~sontag/desz_README.html}.

\section{Conclusions}

In this paper we have defined a new combinatorial measure of redundancy of biological and social networks, 
and have illustrated its efficient computation on several small and large networks. We also noted some 
interesting hypotheses that one could draw from these results such as:
\begin{itemize}
\item
Transcriptional networks are likely to be less redundant than signaling networks. 

\item
The topological redundancy of the {\em C. elegans} metabolic network is largely due to its
inclusion of currency metabolites.

\item
Social networks are prone to be more redundant than biological networks. 

\item
Our calculation of redundancy values and minimal networks
provides a way to gain insight into a predicted orientation 
of a protein-protein-interaction ({\sf PPI}) network 
and determine whether the predicted oriented network has a level of redundancy similar to those in known 
biological networks. 

\item
Our topology-based redundancy measure for biological signaling networks is statistically correlated with 
some measure of the dynamics of the network, namely
higher redundancy is correlated to lower monotonicity and vice versa. 
\end{itemize}
We believe that our fast and accurate computation of redundancy measure will help future researchers
to further fine tune the measure and test it on a large-scale basis.
An interesting question 
that has been partially addressed in the past literature but deserves further investigation is 
to determine the reasons of redundancy of various kinds of biological networks.

\begin{acknowledgments}
We thank Sema Kachalo for the implementation of {\sf NET-SYNTHESIS} and an implementation of 
the random graph generation method of Newman, Strogatz and Watts~\cite{NSW01}.
R\'{e}ka Albert was partially supported by NSF grant CCF-0643529 and Eduardo Sontag was 
supported by NIH grant 1R01GM086881 during this work.
\end{acknowledgments}

\end{document}